\documentclass{PoS}
\usepackage{multicol}
\usepackage{caption}
\usepackage{graphicx}
\usepackage[nosort]{cite}

\newcommand{\hN}{\hat N}
\newcommand{\be}{\begin{equation}}
\newcommand{\ee}{\end{equation}}
\newcommand{\lef}{l_{\rm eff}}
\title{Prospects for large N gauge theories on the lattice}

\ShortTitle{Prospects for large N gauge theories on the lattice}

\author{\speaker{Margarita Garc\'{\i}a P\'erez}\\%\thanks{A footnote may follow.}\\
        Instituto de F\'{\i}sica Te\'orica UAM-CSIC, Nicol\'as Cabrera 13-15, Campus de Cantoblanco, 28049 Madrid, Spain\\
        E-mail: \email{margarita.garcia@uam.es}}

\abstract{
          I will review recent progress on addressing large $N$ gauge theories on the lattice. The focus will be put on the use of large $N$ volume independence as an effective tool to compute non-perturbative dynamics at, otherwise unreachable, large number of colours. A selection of results will be presented and future prospects and challenges for the study of large $N$ QCD and various extensions will be discussed.}

\FullConference{37th International Symposium on Lattice Field Theory - Lattice2019\\
		16-22 June 2019\\
		Wuhan, China}

\begin{document}

\section{Introduction}

\begin{figure}[t]
\begin{minipage}[b]{.5\textwidth}
\centering
\captionsetup{width=.92\linewidth}
     \includegraphics[width=0.91\textwidth]{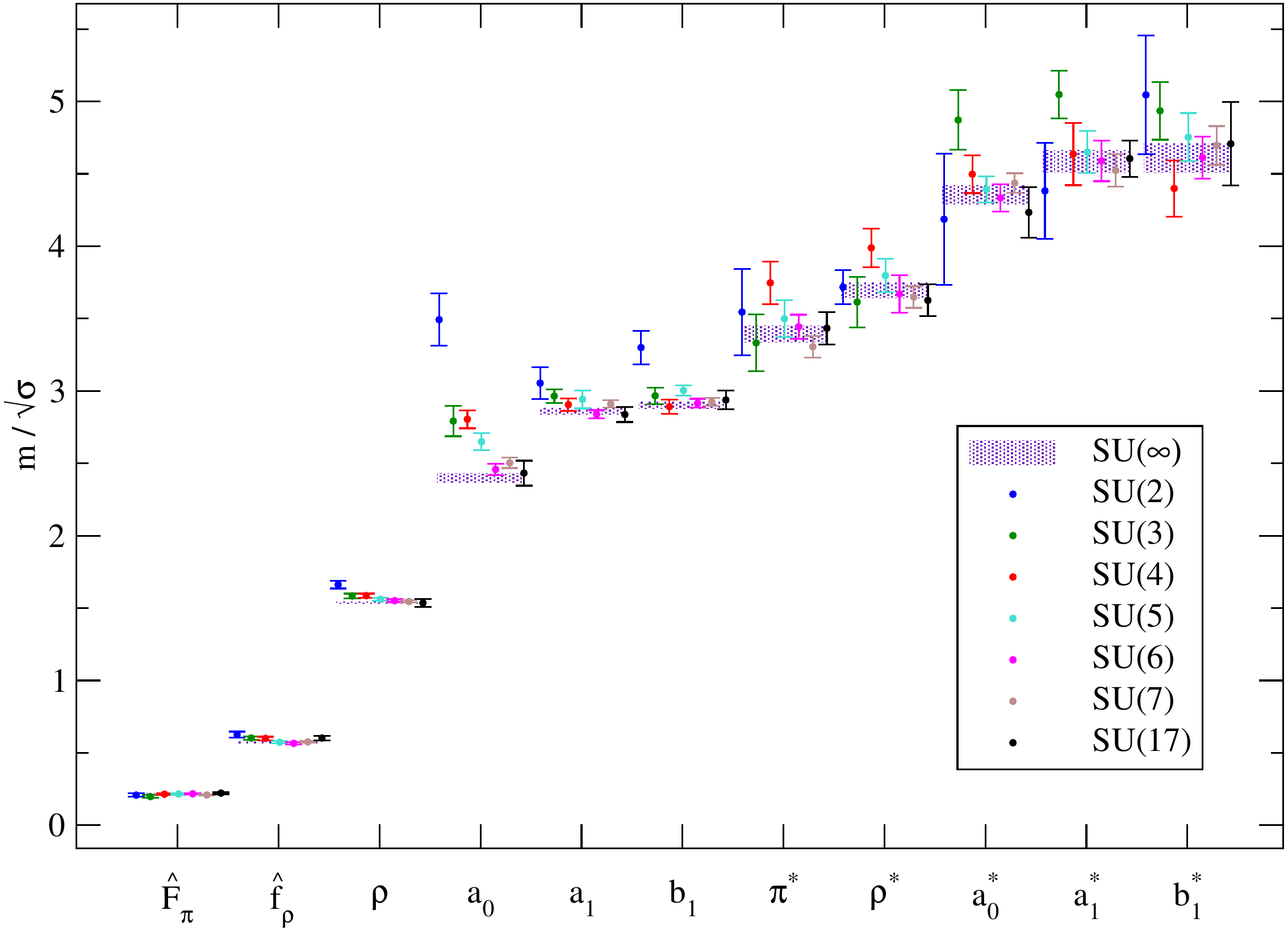}
      \caption{Meson masses and decay constants from~\cite{Bali:2013kia} at a fixed value of the
lattice spacing $a \sqrt{\sigma}=0.2093$.}
     \label{fig1}
\end{minipage}%
\begin{minipage}[b]{.5\textwidth}
\centering
\captionsetup{width=.92\linewidth}
      \includegraphics[width=0.89\textwidth]{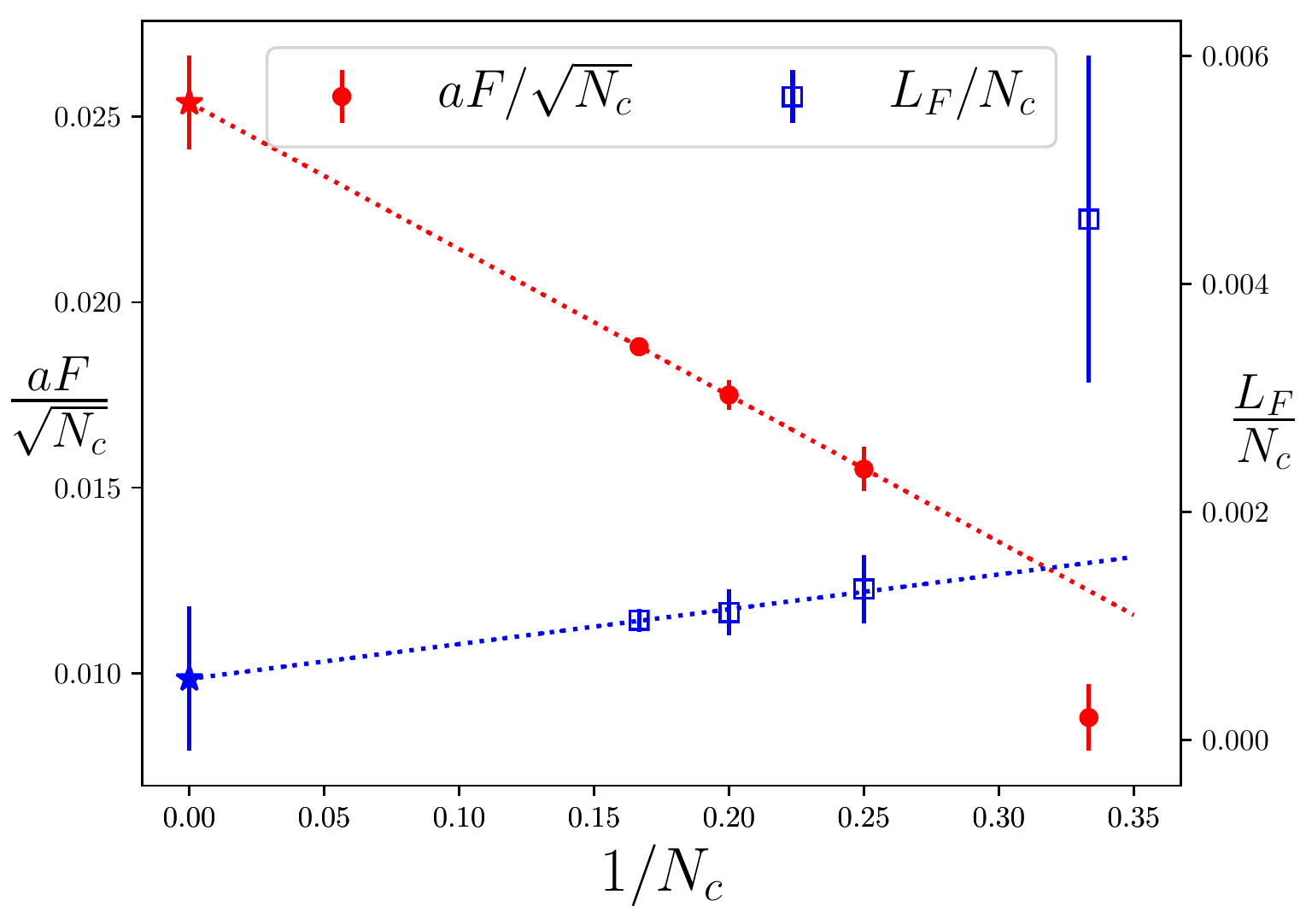}
     \caption{ Scaling with the number of colours of the low energy constants of the chiral Lagrangian $F$ and $L_F$   \cite{Romero-Lopez:2019gqt}.}
     \label{fig2}
\end{minipage}
\end{figure}

Many years have past since 't Hooft pointed out that QCD becomes simpler in the limit of large number of colours, yet we
are still very far from solving the theory in this limit. 
Much like in the real world, non-perturbative quantities in large $N$ QCD  can only be computed {\it ab initio} by numerical lattice simulations. 
It is in this respect that the large $N$ limit seems to offer no particular simplification.
The lattice approach is very demanding for large $N$; in addition to the usual infinite volume and continuum
extrapolations, a quite costly extrapolation in $N$ is required.  As a matter of fact, large $N$ lattice standards are far from those  
in QCD.  The most comprehensive, up-to-date study of the meson spectrum  
is summarized in fig.~\ref{fig1}~\cite{Bali:2013kia}. The results are extracted from quenched simulations with $N\!=\!2$ to 17, 
at one value of the lattice spacing, $a \sqrt{\sigma}=0.2093$.
However, a detailed analysis in  $SU(7)$ shows significant lattice artefacts at this value of $a$~\cite{Bali:2013fya}.
In addition, the effect of quenching can also be 
important, affecting subleading in $N$ corrections that are currently being studied~\cite{DeGrand:2016pur,Donini:2016lwz}  
-- fig.~\ref{fig2} illustrates for instance the scaling with $N$ of some 
low energy constants of the chiral Lagrangian with four active flavours, presented by 
F. Romero-L\'opez at this conference ~\cite{Romero-Lopez:2019gqt}.
All in all, large $N$ lattice simulations are costly and restricted to rather limited values of $N$. Although the dependence in the number of colours seems to be mild, it is convenient to have an approach that allows a more constrained large $N$ extrapolation.

In this regard, it was pointed out long ago by Eguchi and Kawai that large $N$ comes with a remarkable property:
under certain conditions, to be discussed later on, volume effects disappear in the large $N$ limit~\cite{Eguchi:1982nm}. From the practical point of 
view, this is a huge simplification; reliable large $N$ results can be obtained even on a reduced one-point lattice. 
Although the original proposal put forward by Eguchi and Kawai and some of its variants have been shown 
not to work~\cite{Bhanot:1982sh,Bringoltz:2008av,Ishikawa:2003,Teper:2006sp,Azeyanagi:2007su}, 
the idea has been revived in the past few years. 
In this contribution, after reviewing some of the working prescriptions, I will illustrate with several examples the efficiency of this approach
for directly exploring very large values of $N$ (up to $N=841$ for computing the $SU(N)$ string tension~\cite{GonzalezArroyo:2012fx}). 
From the theoretical point of view, the idea is also very powerful; one of the few exact results in gauge theories with direct 
links to non-commutative gauge theory, supersymmetry and string theory, some of which will be highlighted later on. 
Since the focus of the presentation will be on this topic, I will not review other work in the standard lattice approach to the large $N$ limit.
To keep up with the progress in the field, the reader is addressed to the contributions to these 
proceedings~\cite{Romero-Lopez:2019gqt,Beane:2018oxh,Bietenholz:2019zzn,Lee:2019zml,Yamanaka:2019gak,Philipsen:2019bzj,Philipsen:2019rjq,
Hanada:2019rzv,Holligan:2019lma}, several recent works~\cite{DeGrand:2016pur,Donini:2016lwz,Vera:2018lnx,Ce:2016awn,Bonati:2016tvi} and the reviews~\cite{Lucini:2012gg,Panero:2012qx}.

\section{Eguchi-Kawai reduction}

As already mentioned, the essence of the proposal by Eguchi and Kawai is the observation that finite volume effects are absent in the planar large N limit. This statement is exact on 
the lattice provided center symmetry and translational invariance are preserved. 
The proof is non-perturbative and relies on the equivalence of the Schwinger Dyson
equations of Wilson loop expectation values on infinite and finite lattices~\cite{Eguchi:1982nm}. 
The equivalence, known as EK reduction, can be taken to the extreme, reducing the lattice $SU(\infty)$ Yang-Mills theory to a matrix model
defined on a single site. 

Center symmetry is the crucial point for the validity of reduction, since in several instances it is spontaneously broken.  
In fact, it is broken on a 4-dimensional lattice with periodic boundary conditions for the gauge fields, as in the original 
Eguchi-Kawai proposal~\cite{Bhanot:1982sh}.
At weak coupling, it is simple to argue why this should be the case. Consider for instance a finite-temperature set-up in which 
only one direction is reduced by compactifying it on a small circle. 
It is well know that center symmetry is spontaneously broken at high temperature in pure Yang-Mills theory. 
The breaking is manifest at weak coupling in the fact that 
the one-loop effective potential for the Polyakov loop ($P$) is minimized when $P \in \mathbf Z_N \mathbf I$.  
The same happens when all 4 directions are compactified with periodic boundary conditions, the $\mathbf{Z}_N^4$ symmetry is broken at infinite $N$ by quantum fluctuations~\cite{Bhanot:1982sh}.

The thermal analogy brings about one of the ideas to salvage volume reduction; it goes under the name of continuum reduction~\cite{Narayanan:2003fc,Kiskis:2003rd} and
enforces center symmetry by working on lattices with $L$ sites satisfying $L a\! > \!1/T_c$, with $T_c$ the critical temperature for deconfinement. 
Under that condition, volume effects disappear in the large $N$ limit. 
Nevertheless, full reduction can be attained in other approaches that include the use of twisted boundary conditions (TBC)
~\cite{GonzalezArroyo:1982ub,GonzalezArroyo:1982hz,GonzalezArroyo:2010ss}, the addition of adjoint fermions (QCD(Adj))~\cite{Kovtun:2007py}, 
or the use of trace deformed  gauge actions~\cite{Unsal:2008ch}. 
Along this review, I will focus on discussing TBC and QCD(Adj) but, before doing that, I will present two examples of the other two
alternatives at work.

\begin{figure}[t]
\begin{minipage}[b]{.5\textwidth}
\centering
\captionsetup{width=.92\linewidth}
     \includegraphics[width=0.82\textwidth]{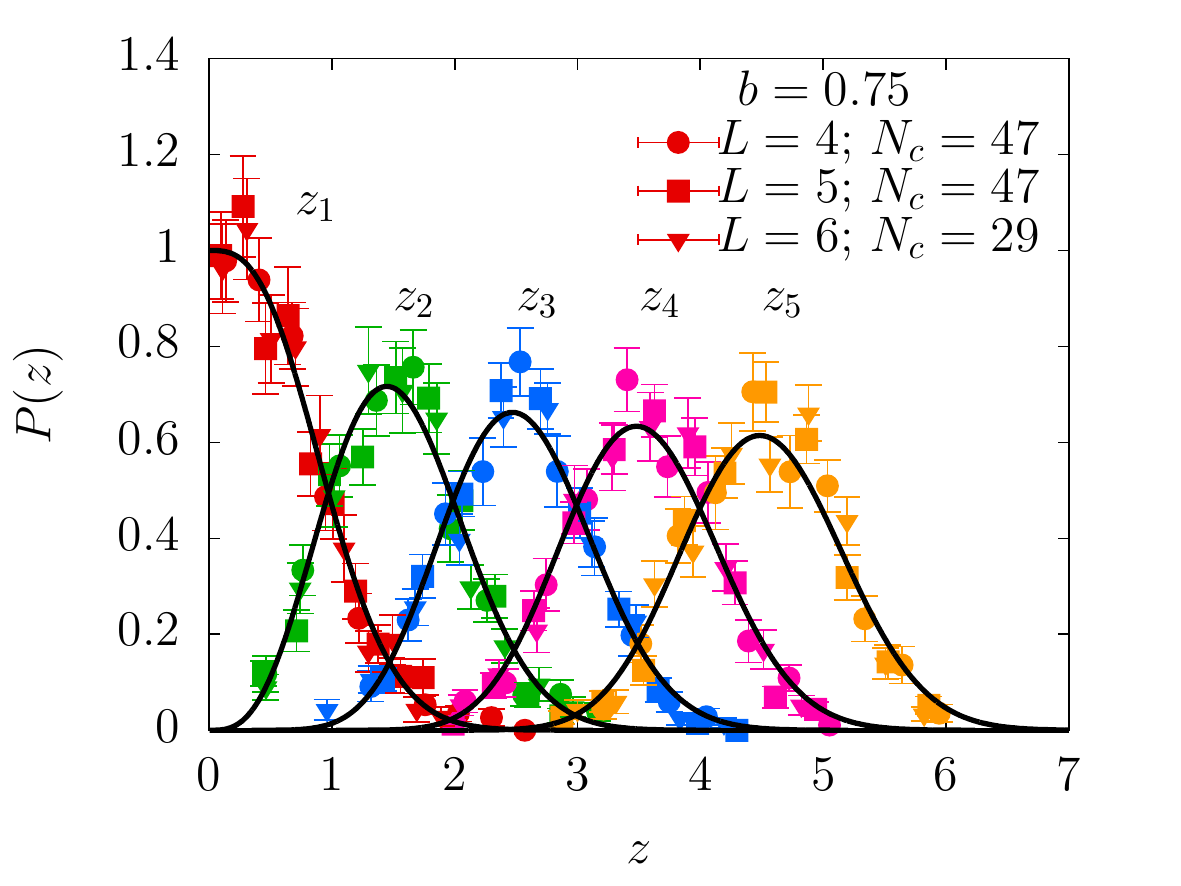}
      \caption{The distribution at $b = 0.75$ of the lowest eigenvalues of the overlap Dirac operator is compared with the prediction from the non-chiral
RMM ($L$ is the lattice size)~\cite{Karthik:2016bmf}.}
     \label{fig3}
\end{minipage}%
\begin{minipage}[b]{.5\textwidth}
\centering
\captionsetup{width=.92\linewidth}
      \includegraphics[width=1.05\textwidth]{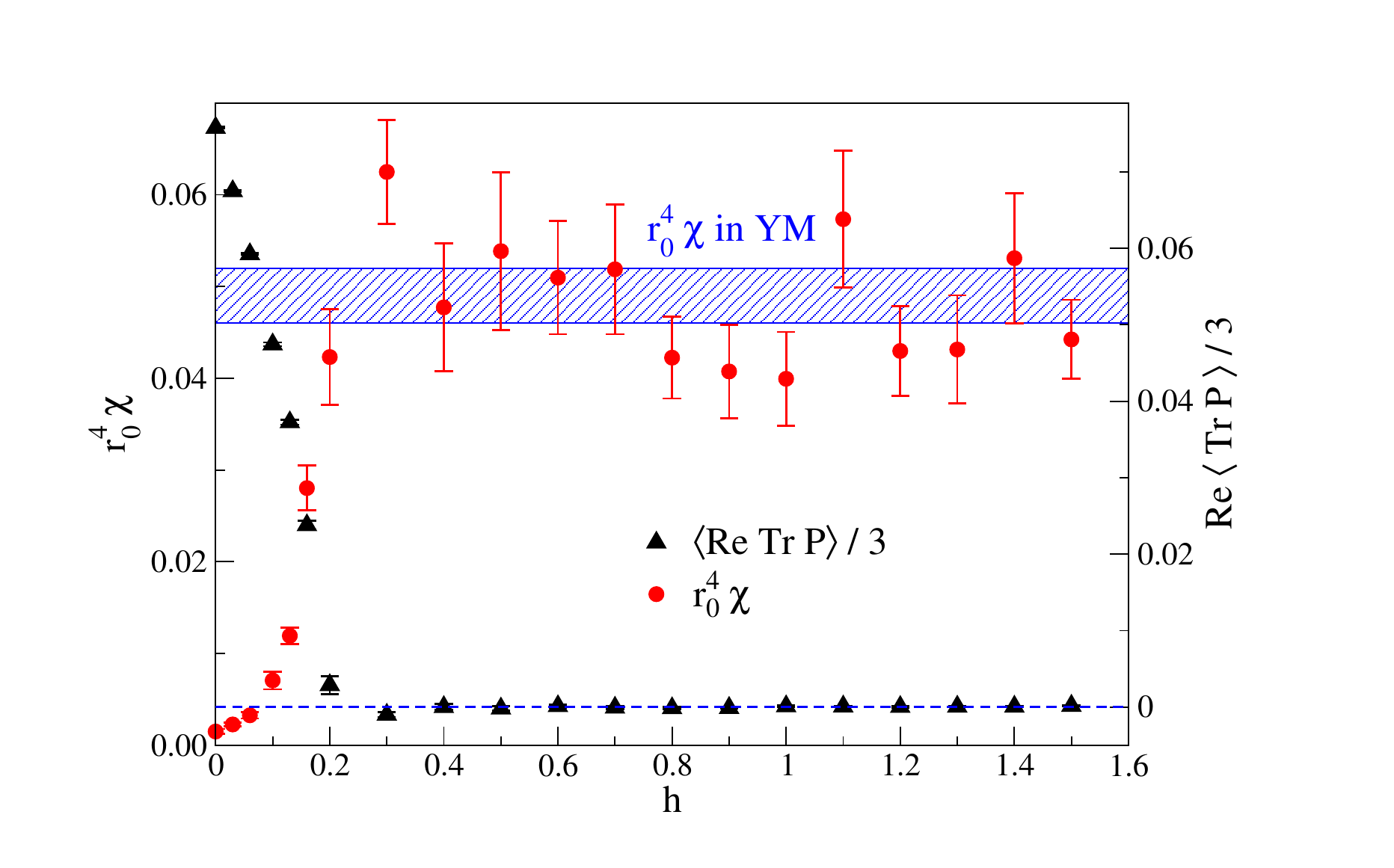}
     \caption{ The topological susceptibility and the expectation value of the Polyakov loop as a function of
the strength of the trace deformation parameter $h$ in eq.~(2.1)~\cite{Bonati:2018rfg}.
}
     \label{fig4}
\end{minipage}
\end{figure}

One of the recent working examples of continuum reduction has been presented in ref.~\cite{Karthik:2016bmf}. It provides a computation of the 
bilinear condensate in 3-dimensional $SU(N)$ gauge theory coupled to $2N_f$ flavours of massless quarks,
for values of $N$ ranging from  7 to 47 on rather small lattices $L=4$, 5, and 6. For an example, fig.~\ref{fig3} displays the lowest lying eigenvalues 
of the overlap Dirac operator, exhibiting no appreciable lattice size dependence and a good comparison with the predictions of a Hermitian non-chiral random matrix model (RMM). The value
of the condensate at $N=\infty$ is estimated to be $\Sigma/\lambda^2 = 0.0042 \pm 0.0004$.

Likewise, M. Cardinali has presented at this conference a study of the $\theta$ dependence in trace deformed $SU(3)$ Yang-Mills theory on $R^3\times S^1$~\cite{Bonati:2018rfg,Bonati:2019unv}. In brief, trace deformations result from the addition to the Yang-Mills action of terms of the form
\be
S_{\rm TD} = h \int d^3 x |{\rm Tr} P(\vec x)|^2
\, ,
\label{eq:traced}
\ee
with $P$ the Polyakov loop along the compact cycle (including also traces of higher powers of $P$  when dealing with larger number of colours). 
This term may be seen as a Lagrange multiplier, enforcing a traceless Polyakov loop even for small compactification radius. The authors of ~\cite{Bonati:2018rfg} have analyzed the dependence of the topological susceptibility and the Polyakov loop expectation value on the strength of the trace 
deformation $h$ -- presented in  fig.~\ref{fig4} for an $8\times 32^3$ lattice at  $\beta=6.4$, which for $h=0$ lies in the deconfinement phase.
Unquestionably, as the parameter $h$ is switched on, both quantities approach the zero-temperature values.
Furthermore, a study in $SU(4)$ has shown that the $\theta$-dependence of the deformed theory coincides with that
at zero temperature whenever there is full center symmetry restoration~\cite{Bonati:2019unv,Bonati:2019kmf}.      

\subsection{Eguchi-Kawai reduction at weak coupling}

Under the use of twisted boundary conditions~\cite{GonzalezArroyo:1982ub,GonzalezArroyo:1982hz} or in  adjoint QCD~\cite{Kovtun:2007py},  the prerequisite that Polyakov loops on the reduced lattice are traceless at weak coupling is satisfied.
In the case of $R^3 \times S^1$, the addition of $N_f$ massless adjoint fermions with periodic boundary conditions in the thermal circle changes the
form of the effective potential for the Polyakov loop to~\cite{Kovtun:2007py,Hosotani:1988bm}
\be
V_{\rm eff} = (2 N_f - 1)\frac{2}{\pi^2 l^4} \sum_{n=1}^{\infty} \frac{1}{n^4} \left|{\rm Tr} \left (P^n\right)\right|^2\, .
\label{eq:veff}
\ee
 Therefore, if the number of flavours is larger that $1/2$ the potential is minimized for ${\rm Tr} \left (P^n\right)=0$. Likewise, the same holds for the QCD(Adj) effective potential on $R^d \times T^n$, computed in~\cite{Barbon:2006us} following     
the techniques put forward in~\cite{Luscher:1982ma,vanBaal:1988qm}.

Comparatively, the starting point with TBC is much simpler.
To be specific, let us discuss the case of a $SU(N)$  pure gauge theory defined on a torus $T^d$ with $d$ even,
period $l$ in all directions and
twisted boundary conditions 
given by~\cite{tHooft:1979rtg,tHooft:1980kjq}: $A_\mu (x + l \hat \nu) = \Gamma_\nu A_\mu (x) \Gamma_\nu^\dagger$, with
 $\Gamma_\nu$ a set of $d$ constant $SU(N)$ matrices subject to the condition:
\be
\Gamma_\mu \Gamma_\nu = Z_{\mu \nu} \Gamma_\nu \Gamma_\mu\, ,  
 \  \  \  {\rm with} \  \  \ 
Z_{\mu \nu} = \exp \left \{ i \epsilon_{\mu \nu} \frac{ 2 \pi k }{\hat N} \right \}\, ,
\ee
where the antisymmetric tensor $\epsilon_{\mu \nu} = 1$ if $\mu< \nu$, and $\hat N = N^{2/d}\in\mathbf{Z}$. If $k$ and $\hat N$ are coprime 
integers, all Polyakov loops with winding number in each direction less than $\hat N$ are traceless on flat connections. 
As a result,  a $\mathbf{Z}_{\hat N}^d$ subgroup of the center symmetry, 
sufficient to guarantee reduction in the large $N$ limit, is preserved at weak coupling.  Still, one would have to show that the remnant center symmetry is maintained non-perturbatively\footnote{Numerical simulations have shown that center symmetry  is broken for the original choice 
of twist with $k=1$~\cite{Ishikawa:2003,Teper:2006sp,Azeyanagi:2007su}. }.
As a matter of fact, this can be achieved if  the flux $k$ is appropriately scaled with  $\hat N$ in a way to be further discussed below~\cite{GonzalezArroyo:2010ss}.

\begin{figure}[t]
\centering
     \includegraphics[width=0.8\textwidth]{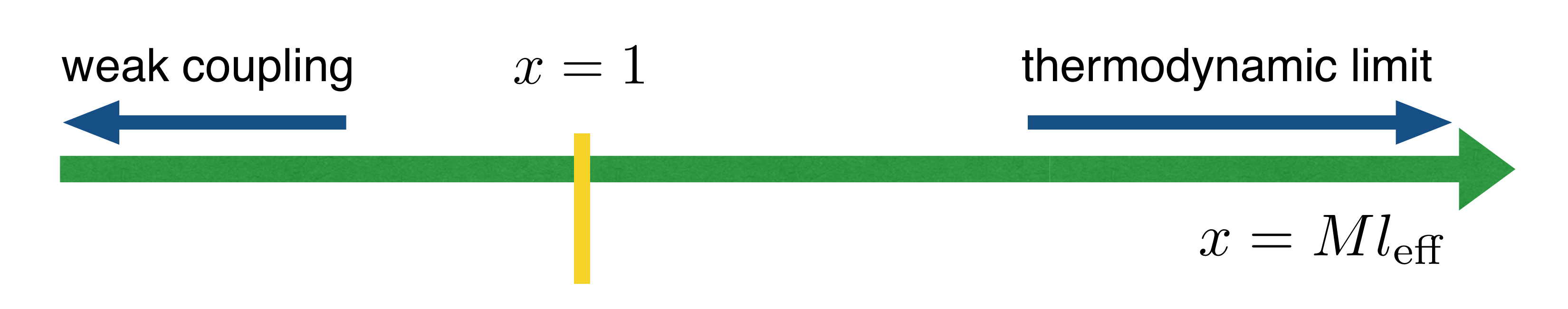}
      \caption{Schematic representation of the relevant scaling variable for volume independence 
$x=M \lef$, where $\lef =\hat N l$ and $M$ represents a relevant mass scale, i.e. $\Lambda_{\rm QCD}$
for 4-dimensional $SU(N)$ Yang-Mills theory. Weak coupling holds at $x\!<\!\!<\!1$ and the strong coupling 
regime sets in at large $x$.}
     \label{fig:game}
\end{figure}

Perturbation theory also reveals how the spatial and colour degrees of freedom are intertwined in these two set-ups;
momentum quantization is in terms of an enlarged effective size: $\lef = \hat N l$~\footnote{$\hat N= N$ for QCD(Adj) on $R^3 \times S^1$. }.  
Notably, the absence of finite volume effects at large $N$ implies that large volume dynamics is akin to large $\lef$, regardless of the concrete value of $l$. 
Furthermore, it leads to conjecture that, even at finite $N$, the relevant scaling variable controlling the dynamics is $\lef$,
or else the dimensionless variable $x= M \lef$, with $M$ a characteristic mass scale of the theory.
It this assumption holds, weak coupling occurs at $x\!<\!\!<\!1$ and the thermodynamic, strongly coupled, regime sets in for large $x$, 
as schematically illustrated in fig.~\ref{fig:game}.

In what follows, I will first review some results derived with the so-called twisted Eguchi-Kawai (TEK) reduction~\cite{GonzalezArroyo:1982ub,GonzalezArroyo:1982hz}, 
in which the full 4 dimensional volume is reduced to a one-site lattice with twisted boundary conditions and
$\lef = a  \sqrt{N}$,  and afterwards I will discuss the use of volume independence 
in a few other contexts, not necessarily involving  large $N$ in the planar limit.

\section{Some large $N$ results obtained from Twisted Eguchi-Kawai reduction}

\begin{figure}[t]
\begin{minipage}[b]{.5\textwidth}
\centering
\captionsetup{width=.92\linewidth}
  \includegraphics[width=0.95\linewidth]{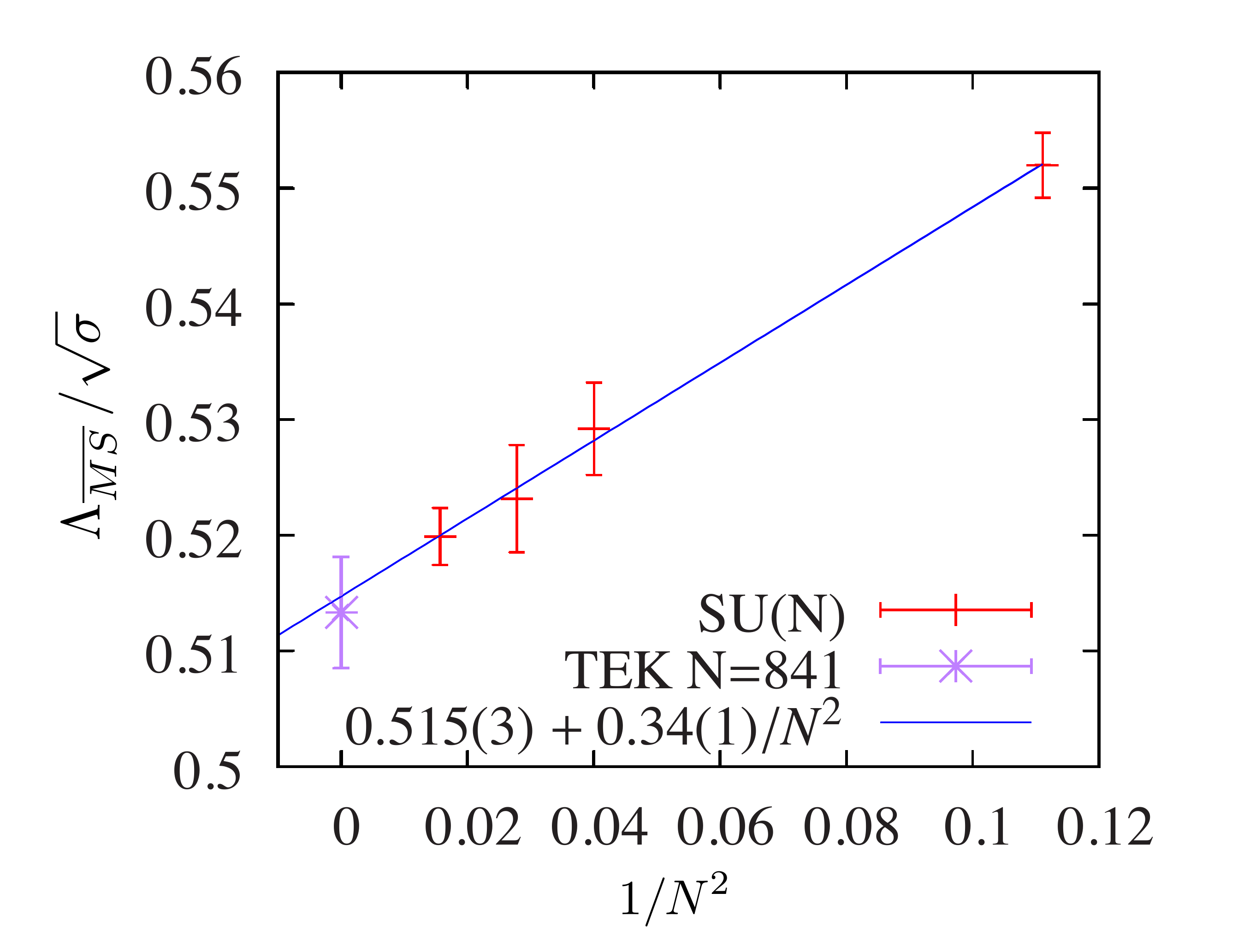}
   \caption{Dependence on $N$ of the $SU(N)$ string tension in the continuum limit~\cite{GonzalezArroyo:2012fx}.}
     \label{fig5}
\end{minipage}%
\begin{minipage}[b]{.5\textwidth}
\centering
\captionsetup{width=.92\linewidth}
   \includegraphics[width=0.95\linewidth]{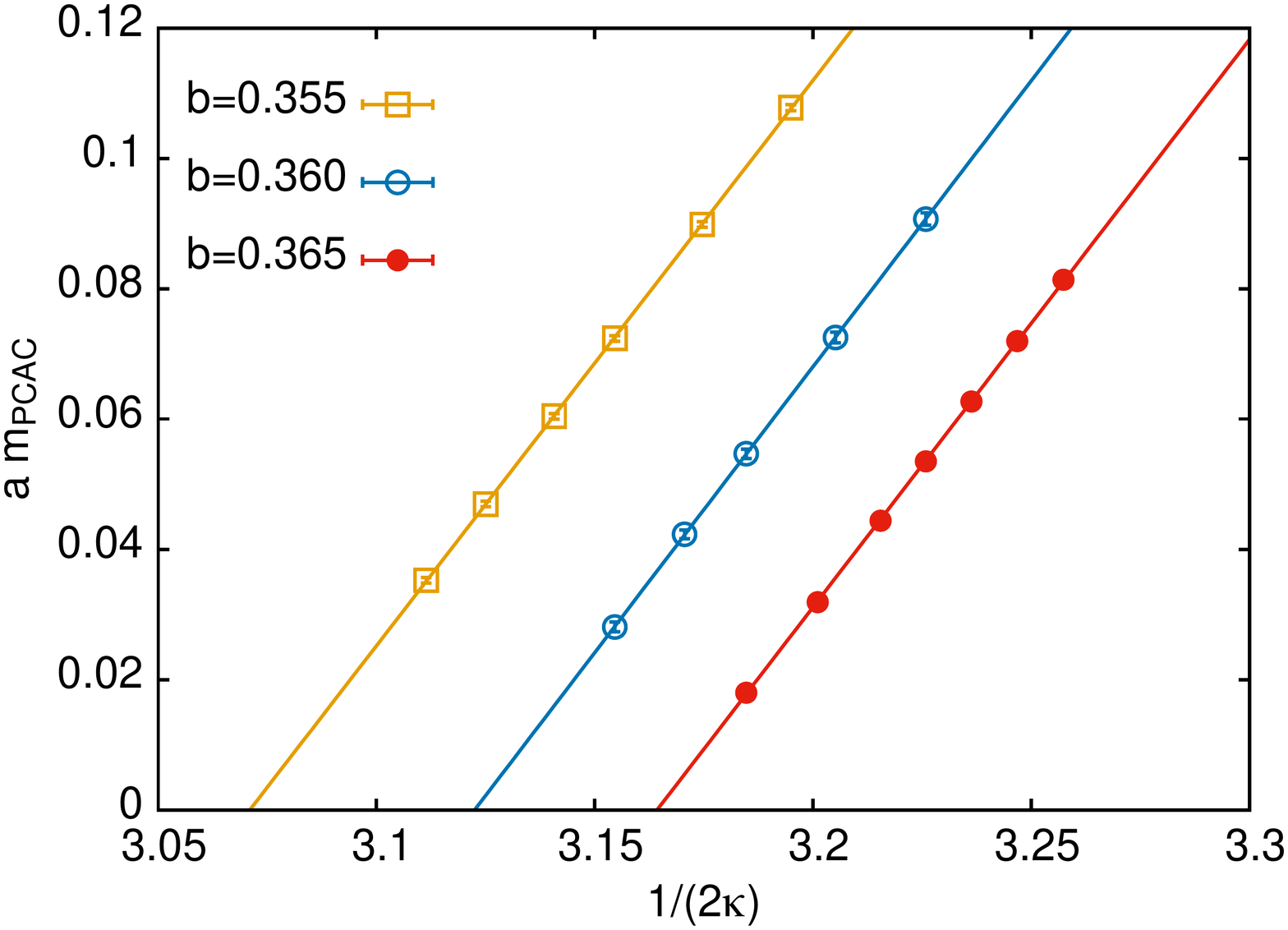}
   \caption{Linear dependence of the PCAC mass on $1/(2\kappa)$ for TEK with $N=289$~\cite{Perez:2020fqn}.}
   \label{fig6}
\end{minipage}
\end{figure}

Twisted Eguchi-Kawai reduction on a one-site lattice has been used in the last few years to study a number of problems in large N gauge theories.
Most of the results correspond to the use of the so-called symmetric twist in 4 dimensions for which the effective size of the 
torus in all four directions scales as: $\lef = a \hat N \equiv a \sqrt{N}$. With this in mind, the large $\lef$ limit is attained 
by sending $N$ to infinity at fixed lattice spacing $a$; this notably leads to the cancellation of non-planar diagrams as advocated for 't Hooft's large N limit~\cite{GonzalezArroyo:1982hz}. 
From a practical point of view, one works at finite $N$, with finite $N$ corrections playing the role of finite volume effects on a 
$\hat N^4$ lattice. As an example of the potential of this approach, fig.~\ref{fig5} displays the inverse of the square root of the string tension in the continuum limit, 
in units of $\Lambda_{\overline{\rm MS}}$, as a function of $1/N^2$~\cite{GonzalezArroyo:2012fx}. The red points correspond to standard lattice simulations
with $N=3,5,6,7$ compared with the result on a one-site lattice with $N=841$. Markedly, the result on the reduced TEK lattice
matches perfectly well the large $N$ extrapolation extracted from the standard results.  

\begin{table}
\begin{minipage}{.5\textwidth}
\centering
\captionsetup{width=.92\linewidth}
  \includegraphics[width=0.9\linewidth]{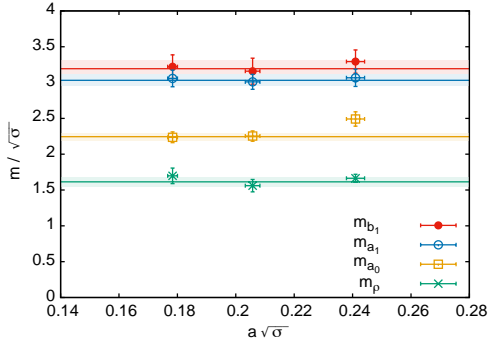}
  \captionof{figure}{Meson masses in the chiral limit as a function of the lattice spacing, 
determined from TEK with $N=289$~\cite{Perez:2020fqn}.}
  \label{fig7}
\end{minipage}%
\begin{minipage}{.5\textwidth}
\centering
\captionsetup{width=.92\linewidth}
\begin{tabular}{c|c|c|}
& $m/\sqrt{\sigma}$~\cite{Perez:2020fqn} & $m/\sqrt{\sigma}$~\cite{Bali:2013kia}\\
\hline 
$\rho$& 1.61(7)(5)  & 1.538(7)\\
\hline 
$a_0$ & 2.24(5)(4)  & 2.40(4)\\
\hline 
$a_1$ & 2.99(8)(2)  & 2.86(2)\\
\hline 
$b_1$ & 3.20(12)(18)& 2.90(2)\\
\hline 
\end{tabular}
\caption{Determination of the large $N$ meson masses in the chiral limit from TEK~\cite{Perez:2020fqn}
and standard lattice simulations~\cite{Bali:2013kia}.}
\label{table1}
\end{minipage}%
\end{table}

Fermions in the fundamental representation can also be introduced in the game. In 't Hooft's large $N$ limit they are quenched and meson propagators can be computed
on the background of gauge configurations generated with a pure gauge numerical simulation. Additionally, one has to deal with the fact that fundamental 
fermions are not compatible with the twist, and we refer the reader to ref.~\cite{Gonzalez-Arroyo:2015bya} to see how to bypass this issue. The basic idea is
that fermions leave on an extended lattice constructed by replicating the background gauge fields. Using this construction, the large $N$ meson spectrum has been computed on
a $\hN^3 \times l_0 \,\hN$ lattice, both with Wilson and twisted mass fermions; some of the results have been presented by A. Gonz\'alez-Arroyo at this conference~\cite{Perez:2020fqn}. As an illustration, we display in fig.~\ref{fig6} the linear dependence of the PCAC mass as a function of $1/(2\kappa)$ 
for $N=289$ at three different values of the lattice spacing.
The lattice spacing dependence of the $\rho$, $a_0$, $a_1$, and $b_1$ meson masses extrapolated to the chiral limit is displayed in fig.~\ref{fig7}, and
the results are compared in table~\ref{table1} with the values obtained through standard lattice simulations in ref.~\cite{Bali:2013kia} (as indicated in the introduction, the latter
have been extrapolated to the infinite $N$ limit but correspond to a fixed value of the lattice spacing $a \sqrt{\sigma}=0.2093$).

Unlike fundamental fermions, adjoint ones are compatible with TBC
and can be directly simulated on a one-site lattice~\cite{Gonzalez-Arroyo:2013bta}. We have analyzed the case of two adjoint Dirac fermions, considered in
the context of walking technicolour theories. The mass anomalous dimension for $SU(289)$ is
$\gamma_* = 0.269 \pm 0.002 \pm 0.05$~\cite{Perez:2015yna}, a value compatible with results derived in standard $SU(2)$ simulations.   
Currently, the mass spectrum is under investigation~\cite{Perez:2020fqn}.

\begin{figure}[t]
\begin{minipage}[b]{.5\textwidth}
\centering
\captionsetup{width=.92\linewidth}
     \includegraphics[width=0.98\textwidth]{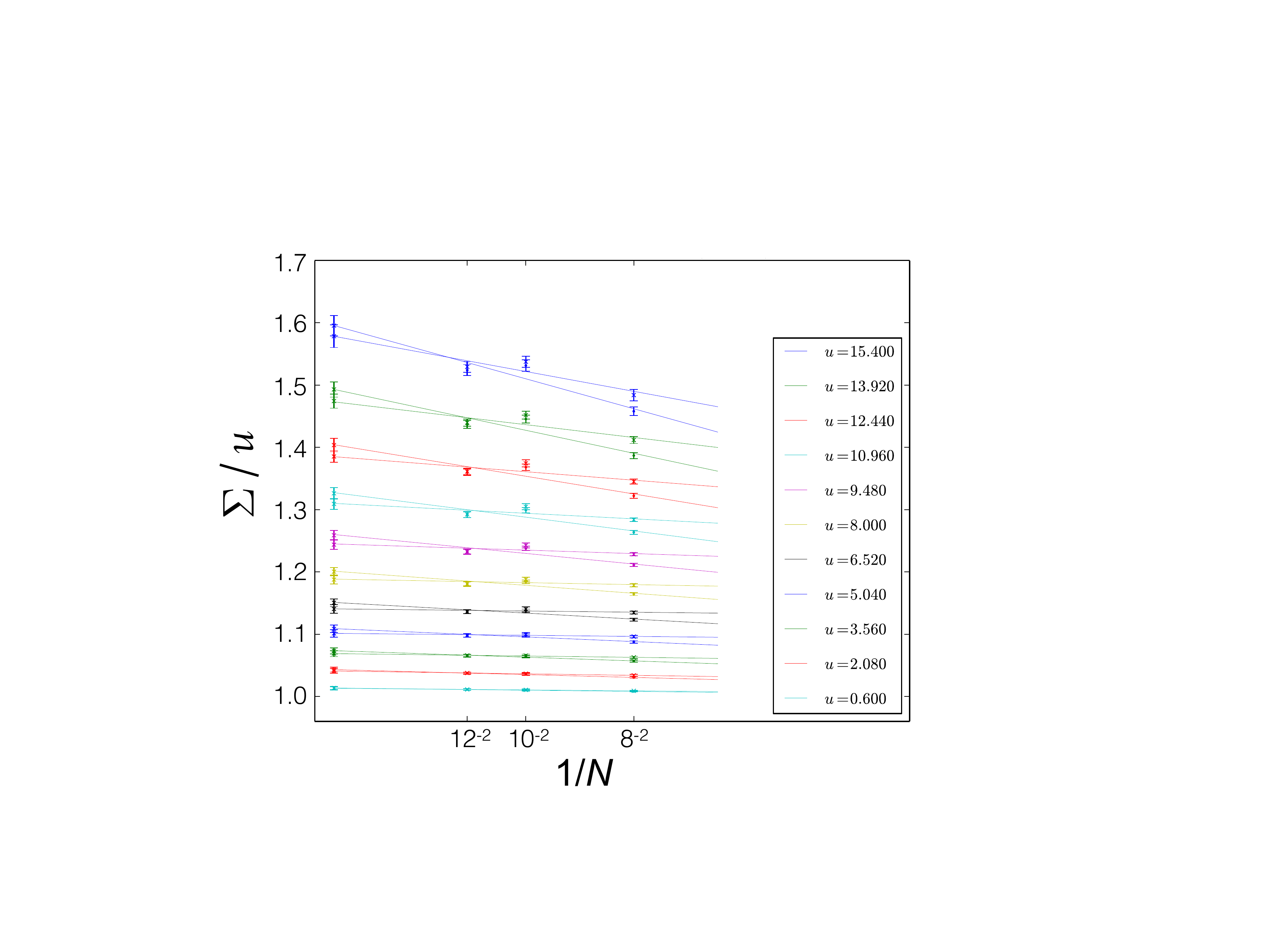}
      \caption{Continuum extrapolation of the step-scaling function $\Sigma/u$ for several values of $u$~\cite{Perez:2014isa}.}
     \label{fig9}
\end{minipage}%
\begin{minipage}[b]{.5\textwidth}
\centering
\captionsetup{width=.92\linewidth}
      \includegraphics[width=\textwidth]{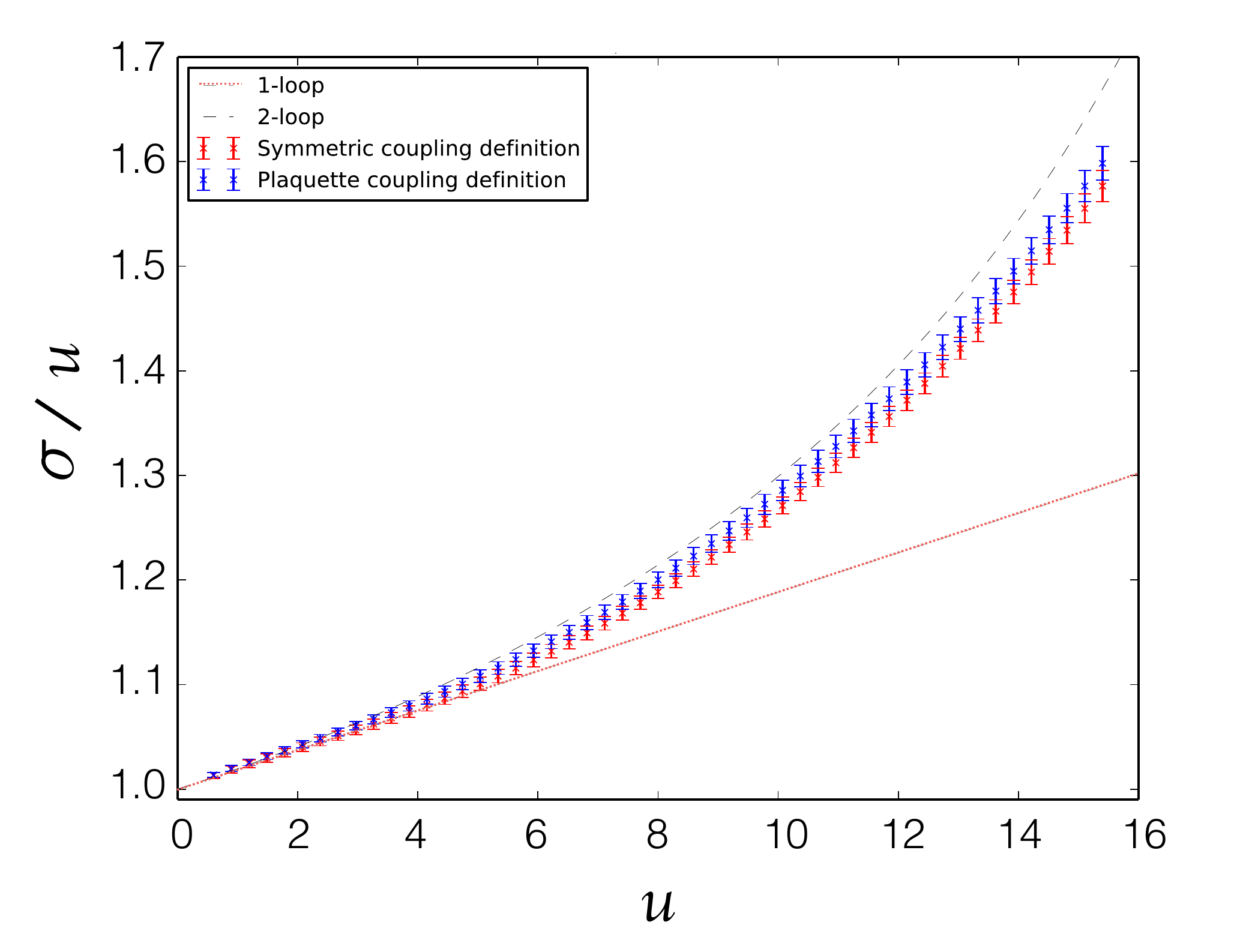}
     \caption{The continuum extrapolated $SU(\infty)$ step-scaling function compared to the one-loop and two-loop
predictions~\cite{Perez:2014isa}.}
     \label{fig10}
\end{minipage}
\end{figure}

The last result I would like to highlight is the determination of the $SU(\infty)$ running coupling in the gradient flow scheme~\cite{Perez:2014isa}. The scale of the coupling is set by  $\lef = a \sqrt{N}$ and step scaling is implemented
by changing $N$, with the continuum limit obtained by sending $N$ to infinity at fixed
value of the renormalized 't Hooft coupling $u\equiv\lambda(\lef)$. Several examples of the continuum extrapolation
of the step scaling function $\Sigma$ are presented in fig.~\ref{fig9};  the continuum extrapolated step scaling function compared with the 1-loop and 2-loop perturbative predictions
is shown in fig.~\ref{fig10}. 

\section{Volume independence at finite and large $N$}

From a theoretical point of view, the idea of volume independence is very powerful, with implications that go beyond those of providing an efficient tool
for the study of large $N$ dynamics. To exemplify this, let us first go back to fig.~\ref{fig:game} and play the standard game of using the volume to explore the dynamics
of the gauge theory. The old dream of the 80s and 90s, trying to link in an analytic way the perturbative, small volume, regime to the confined one~\cite{tHooft:1979rtg,tHooft:1980kjq,GonzalezArroyo:1981vw,Luscher:1982ma,Coste:1985mn,Koller:1985mb,Coste:1986cb,Hansson:1986ia,vanBaal:1986ag,Koller:1987fq,GonzalezArroyo:1988dz,vanBaal:1988va,Daniel:1990iz,GarciaPerez:1993ab,GarciaPerez:1993jw,GonzalezArroyo:1995zy,GonzalezArroyo:1995ex,GonzalezArroyo:1996jp}
~\footnote{See~\cite{vanBaal:2000zc,GonzalezArroyo:1997uj} for a review and further references.}  
goes nowadays under the name of adiabatic continuity -- see i.e. ~\cite{Dunne:2016nmc} -- 
and assumes the absence of phase transitions between
the two regimes.  The hope has always been to extend 
the analytic results at small $x$ to regions where 
non-perturbative phenomena, such as confinement, set in. Specifically, old attempts in SU(2) include for instance the emergence of a mass gap on $T^3 \times R$,
generated, depending on the boundary conditions,  by tunneling through a quantum induced barrier~\cite{vanBaal:1986ag,Koller:1987fq}, or with the contribution of fractional instantons to the perturbatively induced mass gap~\cite{Daniel:1990iz,GarciaPerez:1993ab,GarciaPerez:1993jw,GonzalezArroyo:1995zy,GonzalezArroyo:1995ex,GonzalezArroyo:1996jp}.   
Nowadays, there is a plethora of new $SU(N)$ results in the analytically computable regime ($x\!<\!\!<\!1$), that refer in particular to the case of QCD(Adj) on $R^3\times S^1$. 
A review of results and references can be found in~\cite{Dunne:2016nmc}, they include the analytic computation of the 
mass gap in the pure gauge sector, semiclassically induced by bions~\cite{Unsal:2007jx}, and the spectrum of glueballs, mesons and baryon resonances~\cite{Aitken:2017ayq}.
A large number of results have also been obtained for trace deformed gauge theories, see i.e.~\cite{Dunne:2016nmc} and references therein, and other models such as $CP^N$; for the latter a large $N$ numerical study on $R\times S^1$ has been presented by T. Misumi at this conference~\cite{Misumi:2019upg}. 

\begin{figure}[t]
     \includegraphics[width=1.1\textwidth]{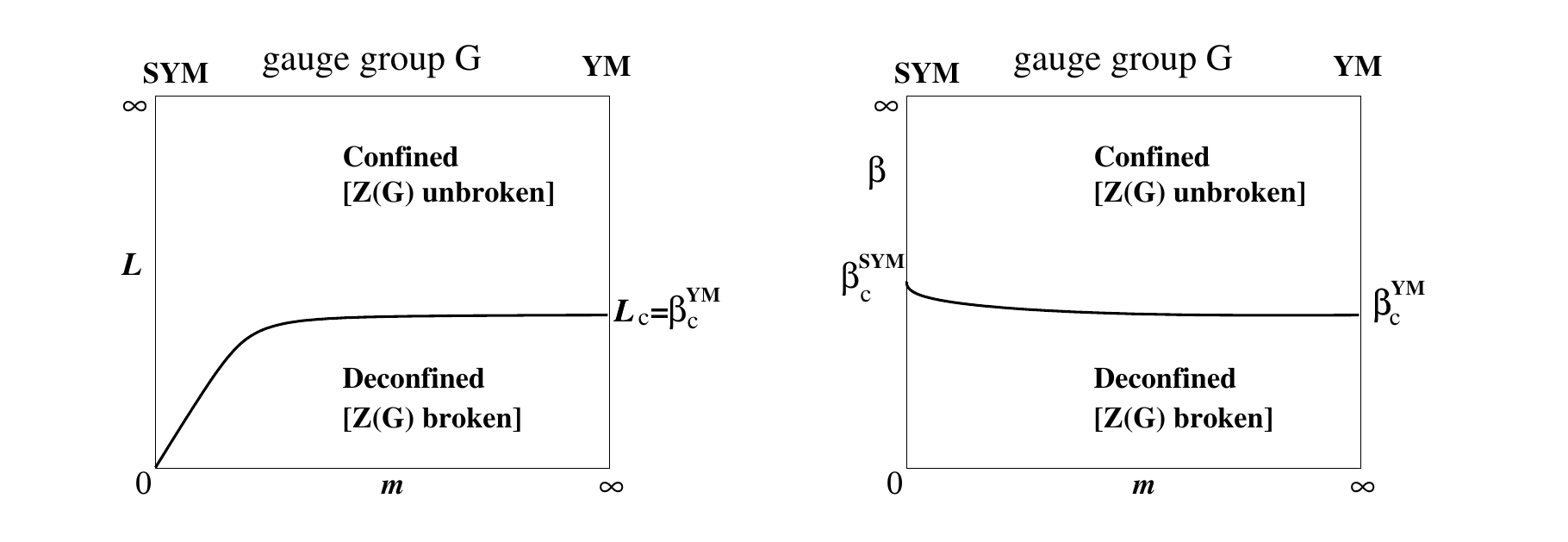}
      \caption{Sketch of the phase diagram of a Yang-Mills theory with an arbitrary simply-connected gauge group $G$
and a single Weyl fermion in the adjoint representation with mass $m$ and periodic (left) and anti-periodic (right) boundary conditions on
$R^3\times S^1$ ~\cite{Poppitz:2012nz}.}
     \label{fig11}
\end{figure}

Another important case of study has been ${\cal N}=1$ supersymmetry on $R^3\times S^1$ deformed away from the supersymmetric case by a
gluino mass term~\cite{Poppitz:2012sw,Poppitz:2012nz,Anber:2014lba}. There is a clear difference in the phase structure of the deformed theory
depending on whether the fermions have periodic, supersymmetry preserving, boundary conditions or thermal ones;
the difference is illustrated for general simply-connected gauge group $G$ in fig.~\ref{fig11}, taken from~\cite{Poppitz:2012nz}.
With pbc and for massless gluinos the theory stays in the $\mathbf{Z}(G)$ unbroken, confined phase, irrespective of the size
of the compact circle. However, as the gluino mass is switched on a phase transition takes place.
The authors of 
ref.~\cite{Poppitz:2012sw,Poppitz:2012nz} argue that one can adiabatically connect the analytically computable phase transition at small $m$ 
with the finite temperature deconfinement phase transition in the pure gauge theory, at infinite gluino mass. This is another instance of adiabatic continuity that can shed light over the (de-)confinement mechanism.      
With this in mind, a lattice investigation of the phase diagram for the case of $SU(2)$ super-Yang-Mills has been presented in~\cite{Bergner:2018unx} and by G. Bergner at this conference. I don't have time to cover in these proceedings the study of the phase diagram of mass deformed non-supersymmetric cases,
the reader is referred for that to the large number of original works on  $S^3\times S^1$~\cite{Hollowood:2009sy}, $R^3\times S^1$~\cite{Cossu:2009sq,Cossu:2013ora}, and $T^4$~\cite{Azeyanagi:2010ne,Bringoltz:2009kb,Bringoltz:2009mi,Catterall:2010gx,Bringoltz:2011by,Koren:2013aya,Lohmayer:2013spa,Cunningham:2013wha}.  

Finally,  I will comment below upon two results that have come up from the combination of adiabatic continuity and large $N$ volume independence: 
emergent fermionic symmetry~\cite{Basar:2013sza}, and singular large $N$ limits in connection to non-commutative
gauge theories~\cite{Perez:2013dra,Chamizo:2016msz,Perez:2018afi}.

\subsection{Emergent fermionic symmetry}

\begin{figure}[t]
     \includegraphics[width=0.9\textwidth]{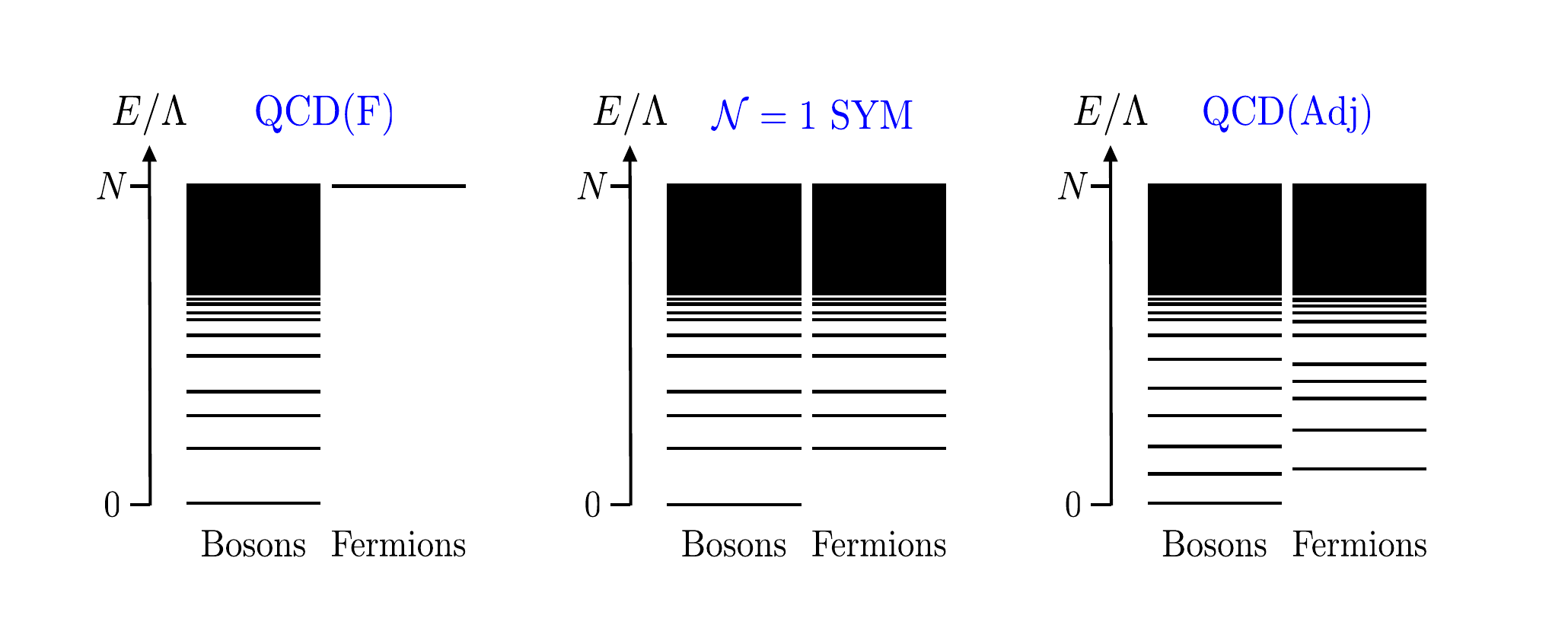}
      \caption{Cartoon of the large $N$ bosonic and fermionic spectrum of: QCD with fundamental fermions (left), ${\cal N}=1$ SUSY Yang-Mills (center), and adjoint QCD with $2N_f >1$ (right)~\cite{Cherman:2018mya}.}
     \label{fig12} 
\end{figure}

Reference~\cite{Basar:2013sza} has pointed out that large $N$ volume independence has necessarily strong implications for the spectrum of Yang-Mills
theory in this limit.   Considering the case of QCD(Adj) on $R^3\times S^1$, the question arises of how to reconcile the expected Hagedorn growth of the
density of states in the large $N$ limit with the fact that the theory becomes independent of the size of the $S^1$ cycle. 
The expected spectrum is schematically depicted in fig.~\ref{fig12}, taken from~\cite{Cherman:2018mya}.
In the supersymmetric case, there is an exact degeneracy between fermionic and bosonic excitations with $E>0$. The natural quantity reflecting this
degeneracy is the Witten index which is independent of the compatification radius $l$.
By analogy, in the non-supersymmetric case one can define a graded partition function:   
\be
I(l) = {\rm Tr} \left [ (-1)^F e^{-l H} \right] = \int dE  \left [ \Omega_B(E) -  \Omega_F(E) \right ] e^{-l E}
\, ,
\ee
with $ \Omega_B$ and $\Omega_F$ the bosonic and fermionic density of states. 
Volume independence implies some restrictions on the dependence on $l$ of this quantity, since physics
becomes independent of $l$ at large $N$. In particular, one expects that the leading Hagedorn growth cancels out in $I(l)$ between fermionic and bosonic
degrees of freedom~\cite{Basar:2013sza}, leading to an emergent fermionic symmetry much alike that of supersymmetry. This proposal connects with similar ideas raised
in the context of planar equivalences at large $N$~\cite{Barbon:2005zj,Hoyos:2005qu} or in non-supersymmetric string theories under 
the name of misaligned supersymmetry~\cite{Dienes:1994np}. Further recent work on this topic can be consulted in~\cite{Cherman:2018mya,Basar:2014jua,Basar:2015asd}. 

\subsection{Singular large $N$ limits, non-commutativity  and the Golden ratio}

We will end up by discussing a type of large $N$ limit that differs from that of 't Hooft in that it preserves non-planar diagrams.
It fits naturally in the discussion that led to fig.~\ref{fig:game} if one considers the double scaling limit in which the number of colours 
is sent to infinity and the size of the torus to zero keeping fixed the value of the effective size and hence $x$. 
This type of limit has been considered first in the context of non-commutative gauge theories~\cite{AlvarezGaume:2001tv,Guralnik:2001ea,Guralnik:2001pv,Griguolo:2001ce,Griguolo:2003kq}. 
It is well known that gauge theories defined on a torus with twisted boundary conditions are equivalent, through Morita duality,
to certain non-commutative gauge theories. As a matter of fact, the Feynman
rules of non-commutative $U(1)$ were for the first time derived from a continuum version of TEK~\cite{GonzalezArroyo:1983ac}, much
before they became fashionable in the string theory literature, see i.e.~\cite{Douglas:2001ba}.  

This connection has been used in the past to argue that gauge theories with TBC are the natural choice for a non-perturbative regulator of
non-commutative gauge theories~\cite{Ambjorn:1999ts,Ambjorn:2000nb,Ambjorn:2000cs}.
In the context of Morita duality, the effective size emerges in a natural way as the size of the
non-commutative torus, and the ratio $\hat \theta = \bar k /\hat N$ ($k \bar k = 1$ (mod $\hat N$)) in the twisted theory 
determines the dimensionless non-commutativity parameter. 
Yet, from this connection only rational values of $\hat \theta$ can be covered. To approach other values, 
ref.~\cite{AlvarezGaume:2001tv} proposed to use a sequence of $SU(N_i)$ theories with $\hat \theta_i = \bar k_i /\hat N_i \rightarrow \hat \theta$. 
However, this limit  is not not guaranteed to be smooth at large $N$; in fact, some
instabilities have been found to occur, first determined in the non-commutative context~\cite{Guralnik:2002ru,Bietenholz:2006cz}.

\begin{figure}[t]
\begin{minipage}[b]{.5\textwidth}
\centering
\captionsetup{width=.92\linewidth}
     \includegraphics[width=0.93\textwidth]{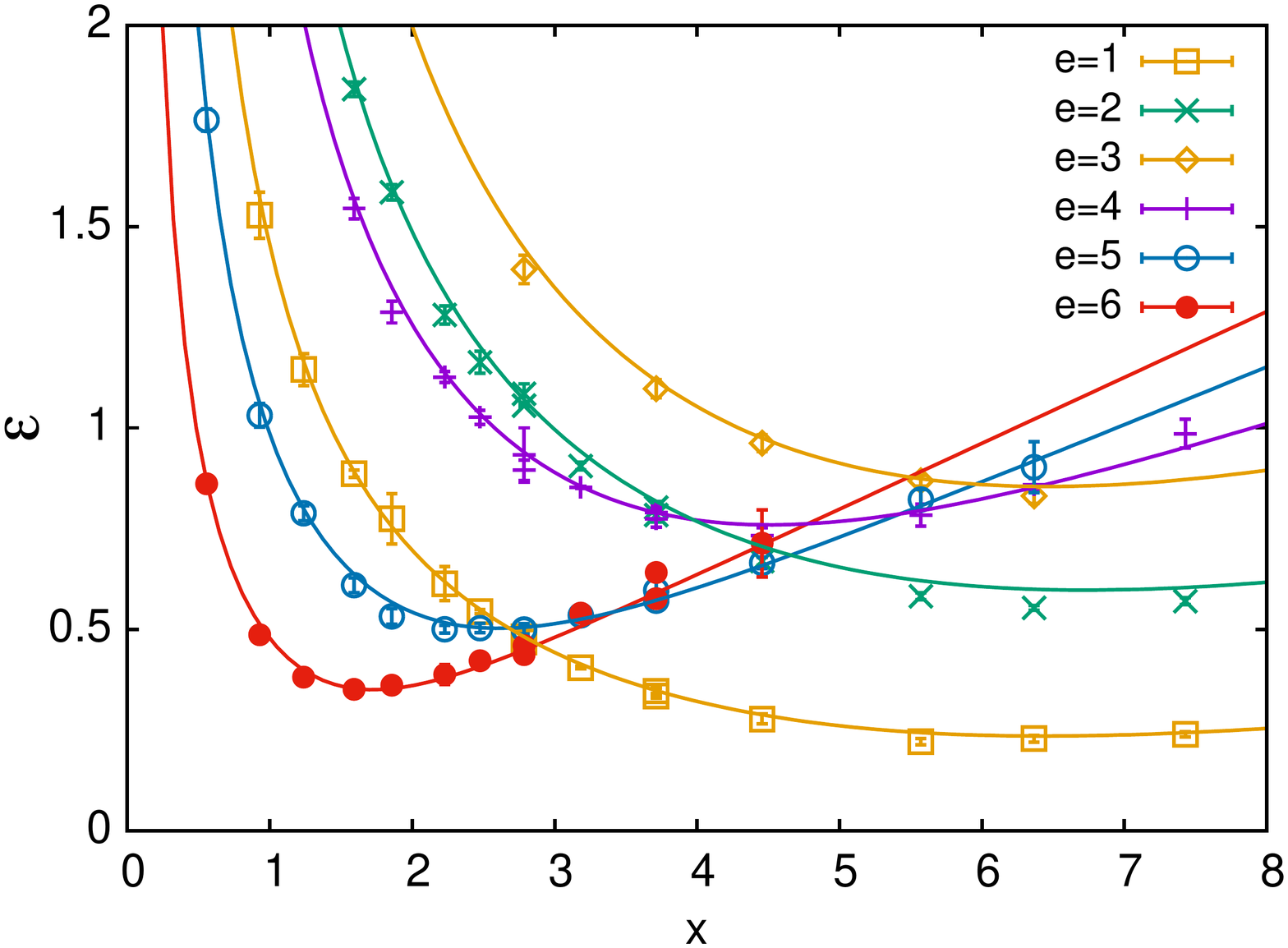}
      \caption{The energy of states with electric flux $(e,0)$ for $N=17$ and $k=3$ compared with the prediction
of eq.~(4.3) with $\chi_0= 0.6$, $\sqrt{\sigma}/\lambda = 0.213$ and $\phi_0(z) = \sin(\pi z)/\pi $~\cite{Perez:2018afi}.}
     \label{fig13}
\end{minipage}%
\begin{minipage}[b]{.5\textwidth}
\centering
\captionsetup{width=.92\linewidth}
      \includegraphics[width=0.98\textwidth]{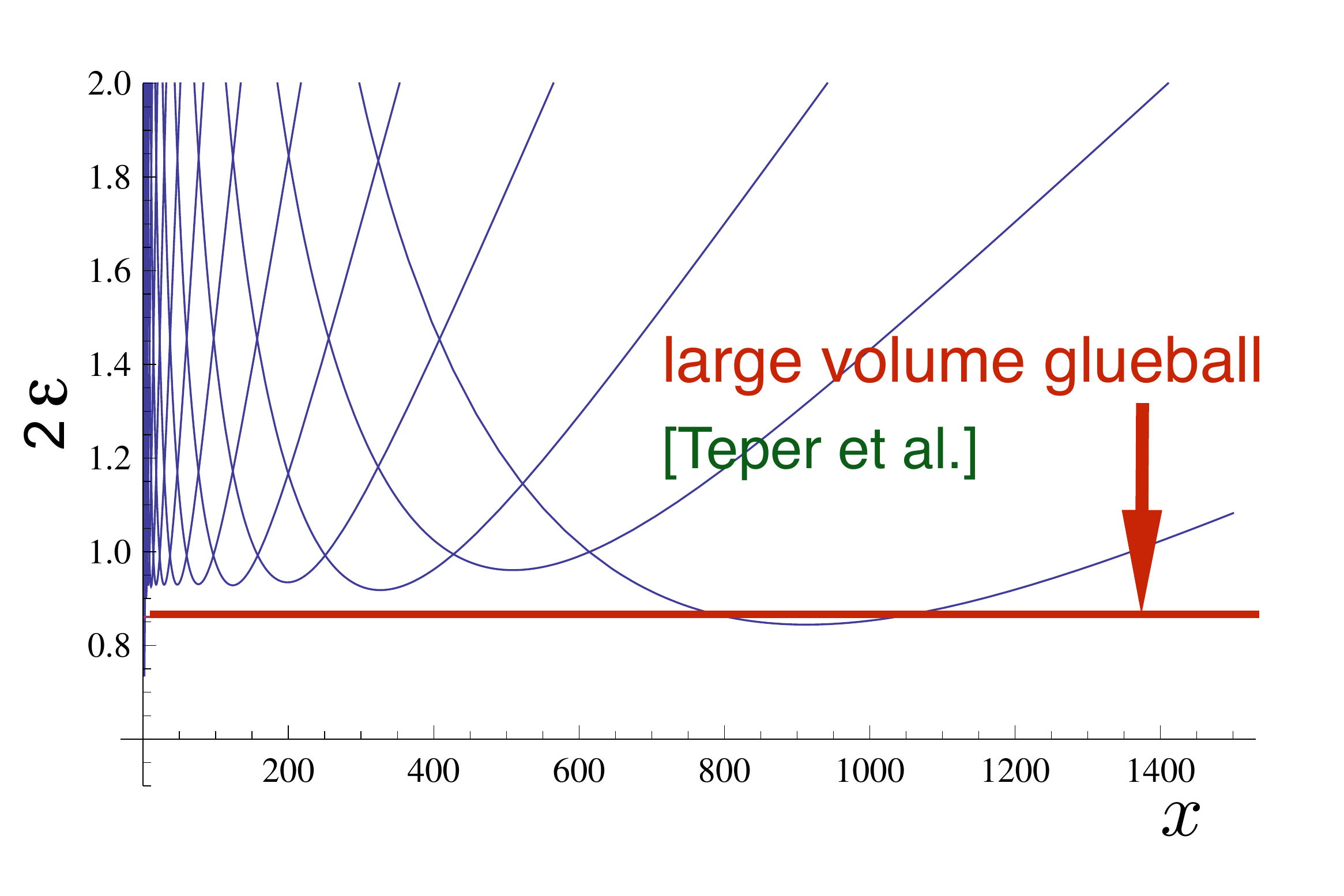}
     \caption{ Twice the energy of electric flux $(e,0)$ with $e$ a Fibonacci number
and $N=1597$, $k=610$. The red line is the value of the glueball mass in $SU(5)$ at $x=7.76$~\cite{Teper:1998te}.}
     \label{fig14}
\end{minipage}
\end{figure}

We have explored in detail this issue for the case of a 2+1 dimensional Yang-Mills theory defined on a twisted $T^2\times R$~\cite{Perez:2013dra,Perez:2018afi}.
The problem appears already at a perturbative level. 
The energy of electric flux with momentum $\vec n$ is found to be at one-loop, in units of the coupling $\lambda$:  
\be
{\cal E}_{\vec n}^2 = \frac{|\vec n|^2}{4 x^2} - \frac{G \left (\hat \theta \vec n \right )}{x} \, .
\ee
The quantity $G(z)$ denotes the self-energy and diverges as $1/||z||$, with
$||z||$ denoting the distance to the closest integer.
When the term $G/x$ dominates, the energy square becomes negative, leading to what is called a tachyonic instability.
Notice that this happens unavoidably if the large $N$ limit is taken at fixed $\bar k$ ($\hat \theta \rightarrow 0$),
as was done in the first TEK prescription~\cite{GonzalezArroyo:1982ub}.
Going beyond perturbation theory, a good description of the energy of electric flux for any value of $x$
is obtained by combining in quadrature the one-loop expression with the expected large $x$ dependence, leading to:
\be
{\cal E}_{\vec n}^2  (x)= \frac{|\vec n|^2}{4 x^2} - \frac{G \left (\hat \theta \vec n \right )}{x} - \frac{\pi \sigma}{3 \lambda^2} \chi_0 + \left ( \frac{4 \pi \sigma}{\lambda^2} \right)^2 
\phi_0^2  \left (  \hat \theta \vec n  \right )  x^2 \, , 
\label{eq:energy}
\ee
where we have taken $\chi_0= 0.6$, $\sqrt{\sigma}/\lambda = 0.213$ and $\phi_0(\vec z)=|(\sin(\pi z_1), \sin(\pi z_2))|/\pi$
(an example of the goodness of the formula for the case of $SU(17)$ with $k=3$ is shown in fig.~\ref{fig13}).

Obviously, tachyonic instabilities can be avoided if ${\cal E}_{\vec n}^2  (x) > 0$, for all values of $x$ and $\vec n$. 
Recently, ref.~\cite{Chamizo:2016msz} has shown that this condition can be translated into one for the quantity:
\be
Z_{\rm min}(N, k) \equiv \min_{m \perp \hat N} m \left |\left | m \hat \theta\right|\right|\, , 
\ee
with $m\perp \hat N$ implying m coprime with $\hat N$.
Choosing $N$ and $k$ such that $Z_{\rm min}(N, k) > 0.1$ is enough to guarantee the absence of instabilities.
Singularly, this quantity seems to be relevant also in  in 4 dimensions,  controlling, for instance, the contribution of non-planar diagrams
to both the expectation value of Wilson loops~\cite{Perez:2017jyq} and the running one-loop 't Hooft coupling in the twisted
gradient flow scheme~\cite{Bribian:2019ybc}.
An open question is whether, for any value of $N$, a flux $k$ can be chosen such that the condition is met. This amounts to
the unsolved Zaremba's conjecture in the mathematical literature. Recently, it has been proven that is is possible for almost all values of $N$~\cite{Huang}. A different issue is whether one can smoothly take 
the singular large $N$ limit described above, leading to any value of the non-commutativity parameter without encountering tachyonic instabilities along the way. 
This turns out to be possible for an uncountable set of irrational values of $\hat \theta$
with, however, zero measure~\cite{Perez:2018afi}.  In particular, one possible sequence corresponds to taking 
$\hat N_i = F_i$ and $k_i= F_{i-2}$, with $F_i$ the ith-number in the Fibonacci sequence, 
leading to a value $\hat \theta = (3 - \sqrt{5}) / 2$~\cite{Chamizo:2016msz}.    
In fact, this choice is  very especial; it might be just a striking coincidence but the lowest  
energy of electric flux, attained for fluxes being themselves in the 
Fibonacci sequence, is very close to half the value of the infinite volume glueball mass computed on standard large volume lattices~\cite{Teper:1998te} (an illustration is presented in fig.~\ref{fig14} for $N=1597$, with energies set by eq.~(\ref{eq:energy})).

\section{Summary}

To sum up, volume independence is a powerful concept that has opened up many interesting avenues. It is an effective tool for
large $N$ lattice simulations, allowing to successfully determine properties of gauge theories in this limit. 
It also encodes relevant information on the nature of these theories even at finite $N$,
encompassing many theoretical ideas: from adiabatic continuity and analytic calculability to emergent fermionic symmetry, 
non-commutative gauge theories and resurgence (a topic not touched in this contribution that has received a lot of attention).  
Overall, I expect many games ahead to be played using this concept and many opportunities to make progress.

\acknowledgments
I would like to thank E. I. Bribi\'an, A. Gonz\'alez-Arroyo, L. Keegan, M. Koren, M. Okawa and A. Ramos for discussions and a fruitful collaboration in  many of these topics, 
and also J. L . F. Barb\'on for discussions. Support from the MINECO/FEDER grant FPA2015-68541-P, the MINECO Center of Excellence Severo 
Ochoa Programme SEV-2016-0597, and the EU H2020-MSCA-ITN-2018-813942 (EuroPLEx) is acknowledged.

%\bibliographystyle{JHEP}

%\providecommand{\href}[2]{#2}\begingroup\raggedright\begin{thebibliography}{100}

%\endgroup

\end{document}